\documentclass{sf2a-conf}
\usepackage[latin1]{inputenc}
\usepackage[comma,authoryear]{natbib}
\usepackage[dvips]{graphicx}

\def\apj{ApJ}%
\def\apjs{ApJS}%
\def\aap{A\&A}%
\def\na{New A}%
\def\pasj{PASJ}%
\def\bbbc{{\mathchoice {\setbox0=\hbox{$\displaystyle\rm C$}\hbox{\hbox
to0pt{\kern0.4\wd0\vrule height0.9\ht0\hss}\box0}}
{\setbox0=\hbox{$\textstyle\rm C$}\hbox{\hbox
to0pt{\kern0.4\wd0\vrule height0.9\ht0\hss}\box0}}
{\setbox0=\hbox{$\scriptstyle\rm C$}\hbox{\hbox
to0pt{\kern0.4\wd0\vrule height0.9\ht0\hss}\box0}}
{\setbox0=\hbox{$\scriptscriptstyle\rm C$}\hbox{\hbox
to0pt{\kern0.4\wd0\vrule height0.9\ht0\hss}\box0}}}}
\def\bbbq{{\mathchoice {\setbox0=\hbox{$\displaystyle\rm
Q$}\hbox{\raise
0.15\ht0\hbox to0pt{\kern0.4\wd0\vrule height0.8\ht0\hss}\box0}}
{\setbox0=\hbox{$\textstyle\rm Q$}\hbox{\raise
0.15\ht0\hbox to0pt{\kern0.4\wd0\vrule height0.8\ht0\hss}\box0}}
{\setbox0=\hbox{$\scriptstyle\rm Q$}\hbox{\raise
0.15\ht0\hbox to0pt{\kern0.4\wd0\vrule height0.7\ht0\hss}\box0}}
{\setbox0=\hbox{$\scriptscriptstyle\rm Q$}\hbox{\raise
0.15\ht0\hbox to0pt{\kern0.4\wd0\vrule height0.7\ht0\hss}\box0}}}}
\def\bbbt{{\mathchoice {\setbox0=\hbox{$\displaystyle\rm
T$}\hbox{\hbox to0pt{\kern0.3\wd0\vrule height0.9\ht0\hss}\box0}}
{\setbox0=\hbox{$\textstyle\rm T$}\hbox{\hbox
to0pt{\kern0.3\wd0\vrule height0.9\ht0\hss}\box0}}
{\setbox0=\hbox{$\scriptstyle\rm T$}\hbox{\hbox
to0pt{\kern0.3\wd0\vrule height0.9\ht0\hss}\box0}}
{\setbox0=\hbox{$\scriptscriptstyle\rm T$}\hbox{\hbox
to0pt{\kern0.3\wd0\vrule height0.9\ht0\hss}\box0}}}}
\def\bbbz{{\mathchoice {\hbox{$\sf\textstyle Z\kern-0.4em Z$}}
{\hbox{$\sf\textstyle Z\kern-0.4em Z$}}
{\hbox{$\sf\scriptstyle Z\kern-0.3em Z$}}
{\hbox{$\sf\scriptscriptstyle Z\kern-0.2em Z$}}}}
\newcommand{\BVF}{Brunt-V\"ais\"al\"a frequency\ }

\newcommand{\beq}{\begin{equation}}
\newcommand{\beqa}{\begin{eqnarray*}}
\newcommand{\beqan}{\begin{eqnarray}}
\newcommand{\greq}{\begin{equation}\left\{ \begin{array}{l}}
\newcommand{\egreq}{\end{array}\right. \end{equation}}
\newcommand{\nngreq}{\[\left\{ \begin{array}{l}}
\newcommand{\nnegreq}{\end{array}\right. \]}
\newcommand{\egreqn}[1]{\end{array}\right. \label{#1}\end{equation}}
\newcommand{\eeq}{\end{equation}} 
\newcommand{\eeqn}[1]{\label{#1}\end{equation}} 
\newcommand{\eeqa}{\end{eqnarray*}}
\newcommand{\eeqan}[1]{\label{#1}\end{eqnarray}}

\renewcommand{\na}{ \vec{\nabla} }

\newcommand{\intvol}{ \int_{(V)}\! }

\newcommand{\vF}{\vec{F}}

\newcommand{\vv}{\vec{v}}

\newcommand{\vx}{\vec{x}}

\newcommand{\demi}{\frac{1}{2}}

\def\Div{\mathop{\hbox{div}}\nolimits}

\begin{document}
\TitreGlobal{SF2A 2006}

\title{Modeling rapidly rotating stars}

\author{M. Rieutord}
\address{Laboratoire d'Astrophysique de
l'observatoire Midi-Pyrénées, UMR 5572, CNRS et Université Paul Sabatier,
14 avenue E. Belin, 31400 Toulouse, France}

\runningtitle{Modeling rapidly rotating stars}

\setcounter{page}{1}
\index{Rieutord, M.}

\date{\today}

\maketitle

\begin{abstract}
We review the quest of modeling rapidly rotating stars during the past
40 years and detail the challenges to be taken up by models facing
new data from interferometry, seismology, spectroscopy... We then
present the progress of the ESTER project aimed at giving a physically
self-consistent model for the structure and evolution of rapidly
rotating stars.
\end{abstract}

\section{Introduction}

The recent observations of rapidly rotating stars with optical and
infra-red interferometers has renewed the interest in modeling
these kind of stars \cite[][]{DVJJA02,Roxburgh04,JMS05}. Indeed,
these data now reveal not only the shape of these stars
\cite[][]{DKJAVFP03} but also the emissivity distribution of the atmosphere
\cite[][]{DKJVONA05,petersonetal06a,petersonetal06b}. With such an
information, models can be adjusted and the inclination of the rotation
axis can be deduced as well as the true equatorial velocity of the star.
Realizing that the brightness of a star depends on the orientation of
its angular momentum vector (equatorial regions are cooler than polar
ones), one wonders what is the real position of rotating stars in the
Herzsprung-Russell diagram, what has been their evolution and therefore
what is their age. The recent discovery \cite[][]{petersonetal06b} that
Vega, a photometric standard, is a rapid rotator seen pole on makes the
foregoing questions all the more exciting.

In this review, I would like to briefly present a short history of
models of rotating stars and then describe the challenges to be taken
up. When I write rotating stars I assume that rotation is not
negligible: I therefore drop the term 'rapidly' since the classification
of a star in the category of rapid rotators is essentially a
matter of precision ( for instance, very precise measurements of frequency of
oscillations require the use of non-perturbative models at much slower
rotations than less resolved data, see \citealt{RLR06}).  All approaches
based on asymptotically slow rotations will not be considered.
I will then give a brief description of the physical phenomena that
should be taken into account when modeling rotating stars; the ESTER
project which aims at forecasting the evolution of rotating stars is
then presented. A brief discussion concludes this short review.

\section{A brief history of models}

The quest of models describing rotating stars up to the break-up limit
(i.e. when the equatorial velocity reaches the keplerian one), really
started with the work of \cite{James64} who calculated the structure of
self-gravitating polytropes at any rotation rate in the permitted range.

This pioneering work was rapidly followed by an attempt from
\cite{RGS65} who went beyond polytropic models using a partition of the
star in two regions: the internal one where centrifugal effect was
assumed small and the external where the Roche approximation could be
used (no self-gravity). Some simple microphysics could be taken into
account (nuclear heating and Kramers type opacities). But with such an
approach the rotation rate was limited and the radiative zone assumed in
a hydrostatic state (which is not possible physically, see below).

The next step was undertaken by an American team (P. Bodenheimer,
J.~Ostrikers and collaborators) who published a series of 8
papers which introduced and exploited the Self-Consistent Field method
\cite[][]{OM68,OB68,Mark68,OH68,Jackson70,BO70,1971ApJ...167..153B,1973ApJ...180..159B}.
The idea of this method is to use the solution of Poisson's equation
(for the gravitational potential $\phi$) expressed with the Green
function, i.e.

\[ \phi(\vx) = -G\int \frac{\rho(\vx')}{|\vx-\vx'|}d^3\vx'\]
which has the great advantage of including the boundary conditions on
$\phi$ at infinity. Again, all dynamics was sacrificed to investigate the
gross role of rotation on bulk quantities of a star (e.g. luminosity,
life time, ...). But the authors faced many difficulties, the code being
not flexible enough to deal with masses less than 9~M$_\odot$ or with very
rapid rotation. Precision of the calculations
was evaluated by the virial test (see sect. 4) and was found to be
4\,10$^{-3}$ \cite[][]{Jackson70}.

Soon after, M. Clement took up the challenge \cite[][]{clem74}. His
method was based on a resolution of the 2D Poisson equation with finite
differences. Compared with preceding ones, results were better at small
masses and problems appeared at large masses. Similarly, the star was
forced to a hydrostatic barotropic state. There too, the virial gave an
estimate of the precision which was found around 2\,10$^{-4}$.
\cite{clem78} examined the effects of prescribed differential rotation
and, \cite{clem79} boundary conditions with a grey atmosphere (among
others). Lastly,
\cite{clem94} investigated the case of differentially rotating
massive stars.

Meanwhile, the Japanese school lead by Y. Eriguchi started a new attempt
with the aim of computing relativistic configurations. An original
method (called EFGH) was first devised \cite[see][]{Eriguchi78} but was rapidly
abandoned because of its difficulty to cope with discontinuities. A new
method \cite[][]{ES81}, close to the Self-Consistent field one, has
then been used to go on the route towards more realistic models. The
first consideration of a baroclinic star appeared with the work of
\cite{UE94,UE95}; this first attempt introduced some baroclinic flow in
the radiative zone but the neglect of viscosity left some degeneracy in
the model \cite[see][]{rieu06}. Ultimately, \cite{SHEM97} focused on
improving microphysics but relaxed on the dynamics keeping a barotropic
configuration.

Parallel to this work \cite{EM85,EM91} developed a code based on a
mapping of the star where the new radial variable $\zeta$ introduces
surface coordinates similar to the surface of the star

\[ r_i(\theta_k) = \zeta_i R_s(\theta_k)\]
with $\zeta_i$ regularly spaced between 0 and 1.
Discretization is with finite differences radially and angularly. The
code seems to be very robust being able to compute configurations very
far from sphericity but the precision of calculations again tested with
the virial theorem seems to be around $5\,10^{-4}$. In \cite{EM91}, a
meridional flow was computed; however, as the momentum equation was
not solved for this flow, this flow should be considered as a mere
illustration of the thermal disequilibrium brought about by the imposed
barotropicity.

As mentioned in the introduction, most recent work has been published by
\cite{Roxburgh04} and \cite{JMS04,JMS05}. Models are still barotropic but
microphysics is improved as well as the precision of the models.
\cite{JMS05} report a virial test at $10^{-5}$.

The conclusion of 40 years of research on modeling rotating stars is
that still large parts of the problem remain unexplored. None of the
models have reached the state of computing self-consistently a
baroclinic radiative zone, which is unavoidable in a rapidly rotating
star (like Vega or Altair). Nobody has ever considered the case of
evolution under fast rotation except with spherically symmetric model
\cite[][]{MM00}. These are very difficult questions but the recent
observational result of interferometry as well as the forthcoming data
from seismology (COROT satellite) force us to go beyond these obstacles.

\section{The ideal model for rotating stars}

The foregoing review of the attempts made at modeling rotating stars
showed us that models are still quite far from actual stars. The reader
may wonder how far. This question brings us to the description of the
ideal model for a rotating star.

Such a model should describe the mean state of a star at any time of its
life and especially the new quantity specific to these stars: angular
momentum.

Unlike a non-rotating star which is a one-dimensional object (in
a large-scale description) which needs only scalar fields (I forget
magnetic fields), a rotating
star is, at least, a two-dimensional object with, at least, one vector
field in addition to all scalar field. Hence, complexity increases not
only by the multi-dimensional nature of the model but also by the number
of physical quantities to be determined. This implies that the ideal model
deals consistently with angular momentum and especially the losses and
gains through stellar winds and accretion. Such a model should also take
into account the baroclinicity of radiative zones and there, the anisotropic
turbulence which appears through shear instabilities; it should also include a
mean-field theory of convection to forecast Reynolds stresses and heat
flux. Of course, observers would like to know the emissivity of the
atmosphere as a function of latitude (if they use interferometry) or as
a function of wavelength if they do spectroscopy. But if they do
asteroseismology they surely wish to know the eigenspectrum of these
objects.

The foregoing points show that progress in the understanding of rotating
stars needs also some advances in the following questions of stellar physics:
\begin{itemize}
\item How angular momentum is distributed in a star and how is it input
or output with what consequences ?

\item The immediately following question concerns the nature of the
Reynolds stresses in the convective and radiative zones.

\item Then, what is the baroclinic state of the radiative regions ?

\item Similarly, the atmosphere is in a  baroclinic state and cannot
be at rest: how strong are the differential rotation and the meridional
currents? Does the atmosphere develop strong azimuthal winds streaming
around the star like Jupiter's winds?

\item Gravity darkening can be so efficient that equatorial regions are
cool enough to develop convection; this raises the question of the latitude
dependence of emissivity of the atmosphere beyond the Von Zeipel model
\cite[see the attempt of][]{LDS06}.
\item I did not mention magnetic fields. Clearly they multiply the
number of problems and first steps should ignore them if possible.
\end{itemize}

\section{The ESTER project}

Crazily enough, the ESTER project (Evolution STEllaire en Rotation) 
takes up the challenge of at least producing a physically self-consistent
model of a rotating star at any rotation rate and let it evolve.

Technically, we construct a two-dimensional model using coordinates
adapted to the geometry of the star (in fact to all its interfaces).
Thus doing, we follow the work of \cite{BGM98} which uses a
mapping between spheroidal coordinates and spherical coordinates
\cite[e.g.][]{RDLCP05,RLR06}. Such a mapping is necessary to impose
correctly boundary or interface conditions.

Then we discretize the equations using spectral methods both radially,
with Chebyshev polynomials, and horizontally, with spherical harmonics.

As our predecessors, we first controlled our model with polytropes in
solid body rotation. The results have been compared successfully to those of
\cite{James64} and then tested by the virial theorem which demands that
\begin{equation} 2T/W + 3P/W + 1= 0\end{equation}
where

\[ T = \intvol \demi\rho \Omega_*^2r^2\sin^2\theta dV, \qquad
 W = \demi \intvol \rho\phi dV, \qquad P = \intvol P dV\]
are respectively the kinetic, gravitational and internal energy.
Typically, the virial equation is satisfied with a precision better than
$10^{-8}$. These very precise models turned out to be excellent for
asteroseismological purposes \cite[see][]{LRR06,RLR06}.

In the next step we generalized the simple polytropes to stars composed
of many polytropic layers \cite[e.g.][]{RDLCP05}.

However, polytropes are barotropic stars and they cannot describe the
dynamics of radiative zones properly. Real rotating stars have indeed
radiative zones which are the seat of slow baroclinic flows which show
up as a differential rotation, a meridional circulation and certainly
some weak anisotropic turbulence. As such flows were largely unknowns,
we decided to first investigate them, {\it per se}, in some simplified
context. Thus, in \cite{rieu06} we solved this problem in the case
of spherically symmetric star made of a Boussinesq fluid, {\it i.e.}
with a fluid of negligible compressibility.  We could thus determine
how the differential rotation of a radiative zone is controlled by the
baroclinic torque and what was the role of the Ekman boundary layers. We
also showed the dynamical role of the viscosity jump at the core-envelope
boundary; it gives birth to a Stewartson layer parallel to the rotation
axis which reaches surface layers. It was shown to be destroyed by
molecular weight gradients which develop as the star evolves.

The next step has been to overcome the Boussinesq approximation and
consider a more realistic radiative zone. For this purpose we enclosed a
self-gravitating, compressible, viscous, rotating fluid inside a rigid sphere.
Taking into account nuclear energy heating and radiative opacities in
the Kramers form, we solved

\begin {equation}
\label{dim_eq}
\left\{\begin{array}{l}
\displaystyle \Delta\phi=4\pi G\rho\\
\displaystyle \rho T\vec{v}\cdot\nabla s=-\Div\vec{F}+\varepsilon\\
\displaystyle
\rho\left(2\vec{\Omega}\times\vec{v}+\vec{v}\cdot\nabla\vec{v}\right)=
-\nabla p-\rho\nabla\left(\phi-\frac{1}{2}\Omega^2r^2\sin^2\theta\right)
+\mu(\Delta\vv+\frac{1}{3}\na\Div\vv)\\
\displaystyle \Div(\rho\vec{v})=0
\end{array}
\right.
\end{equation}
where we used the microphysics

\[
\varepsilon=\varepsilon_0 X^2\rho^2T^{-2/3} \mathrm{e}^{-bT^{-1/3}},
\qquad \vF=-\frac{16\sigma T^3}{3\kappa\rho}\na T
\]
In order to describe the radiative transport of energy, we use the 
opacity given by a Kramer's law $\kappa=\kappa_0 T^{-\beta} \rho^\eta$

\begin{figure}[t]
\centerline{\includegraphics[width=0.5\linewidth,angle=0]{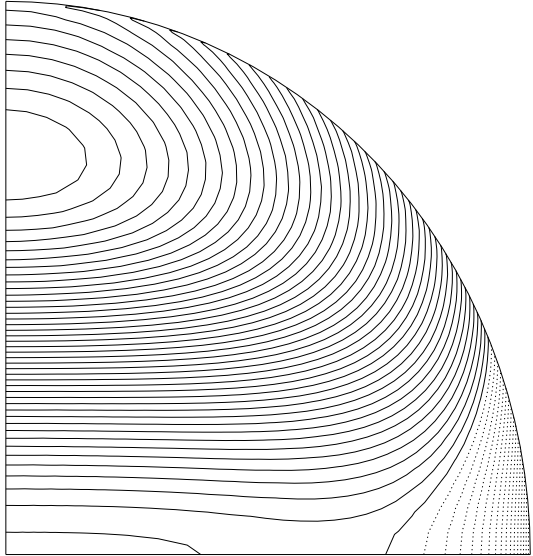}
}
\caption{The Brunt-V\"ais\"al\"a squared in the meridional plane of a
rapidly rotating star: Note the convectively unstable region near the
equator ($N^2<0$).}
\label{ester}
\end{figure}

A detailed description of the results may be found in \cite{ER06}.
We show in Fig.~\ref{ester} a meridian view of the distribution of the
squared \BVF. The remarkable results is the appearance, in the
equatorial region, of convectively unstable fluid ($N^2<0$). A region
which disappears if the rotation is slow enough.

\section{Discussion}

The results of \cite{ER06} show that it is now possible to compute in a
self-consistent manner a rotating radiative zone in a steady state, with
realistic opacities, nuclear reactions and velocity fields. Yet, the
star is confined in a spherical box which supports some pressure. The
next step is of course to relax this constrain and move to spheroidal
geometry. Then it will be possible to apply more realistic boundary
conditions on the temperature or plug a model of atmosphere. However,
before that, we still need to find a good (or not too bad) model for
the convection zones. Such a model should generalize in two dimensions,
the approach of the mixing length theory and give a reasonable account
of the convective flux.

Then, after some calibration tests using one-dimensional models,
more physics will be taken into account. Most difficult steps will
concern all transport phenomena which are related to turbulence. Indeed,
turbulence controls the Reynolds stresses and thereby the diffusion of
angular momentum which is closely related to the diffusion of elements
\cite[][]{zahn92}.

\begin{acknowledgements}
I would like to thank F.~Espinosa and B.~Pichon for interesting
discussions about this wide subject. I also thank the organizers of the
``Journées de la SF2A" for giving me the opportunity to present this
fascinating subject.
\end{acknowledgements}

\end{document}